\documentclass[useAMS,usenatbib,letters]{mn2e}

\usepackage{color}
\usepackage{graphicx}
\usepackage{amsmath}
\usepackage{amssymb}
\definecolor{orange}{rgb}{1,0.5,0}
\usepackage[colorlinks=True,citecolor=orange,urlcolor=blue,linkcolor=red]{hyperref}

\pdfoutput=1

\newcommand{\msun}{{\rm M}_{\sun}}

\def\apj{ApJ}%
\def\apjl{ApJ}%
\def\apjs{ApJS}%
\def\aap{A\&A}%
\def\mnras{MNRAS}%
\begin{document}

\title[MHD models for Orion Source I disk-winds]
{MHD Modeling of a Disk-Wind from a High-Mass Protobinary: the case of Orion Source I}
\author[B. Vaidya, C. Goddi]{B. Vaidya$^{1}$\thanks{E-mail:
B.Vaidya@leeds.ac.uk (BV)}, C. Goddi$^{2,}$$^{3}$ 
\\
$^{1}$School of Physics and Astronomy, University of Leeds, Leeds LS29JT\\
$^{2}$European Southern Observatory, Garching, Germany
$^{3}$Joint Institute for VLBI in Europe, Postbus 2, 7990 AA
Dwingeloo, The Netherlands}

\date{\today}
\pagerange{\pageref{firstpage}--\pageref{lastpage}} \pubyear{yyyy}

\maketitle

\label{firstpage}

\begin{abstract}
Very long baseline interferometry (VLBI) observations of SiO masers in
Orion Source I has enabled for the first time to resolve the outflow from a high-mass protostar in the launch and collimation region. Therefore, Source I provides a unique laboratory to study  mass-loss
and mass-accretion  in a high-mass protostar. 
We numerically simulate the dynamics of the disk-wind inside 100
     AU from Source I.
This enables us to investigate  the balance of different forces
(gravitational, magnetic, thermal) regulating gas dynamics in massive star formation. In this work, we adopt magnetohydrodynamic (MHD) disk-wind models to explain the
      observed properties of the disk-wind from Orion Source I. 
The central source is assumed to be a binary composed of two 10\,$\msun$ stars
     in a circular orbit with an orbital separation of 7 AU. High resolution ideal MHD wind
     launching simulations (which prescribe disk as a boundary) are performed using the PLUTO code. The simulations are allowed to run until a steady
     state is obtained. MHD driven disk-wind provides a consistent model for the wide-angle
  flow from Source I probed by SiO masers,  reproducing  the bipolar
  morphology, the  velocity amplitude and rotational profile, the
  physical conditions, and the magnetic field strength.

\end{abstract}



\section{Introduction} 

In the formation of low-mass stars, disks and jets play a major role in controlling mass  accretion
 and removing excess angular momentum. 
Specifically, low-mass protostellar jets are expected to be launched and collimated via magneto-centrifugal processes at distances $<<$10 AU from Young Stellar Objects or YSOs \citep[e. g.][]{Konigl:2000p607}. 
  However, the balance of forces in proximity to high-mass YSOs has
  been difficult to infer because high extinctions, clustering, and
  large distances hinder attempts to resolve the gas structure and
  dynamics at small radii ($<$100 AU) where outflows are launched and
  collimated from accretion disks.
  
\begin{figure} \centering
\includegraphics[width=0.8\columnwidth]{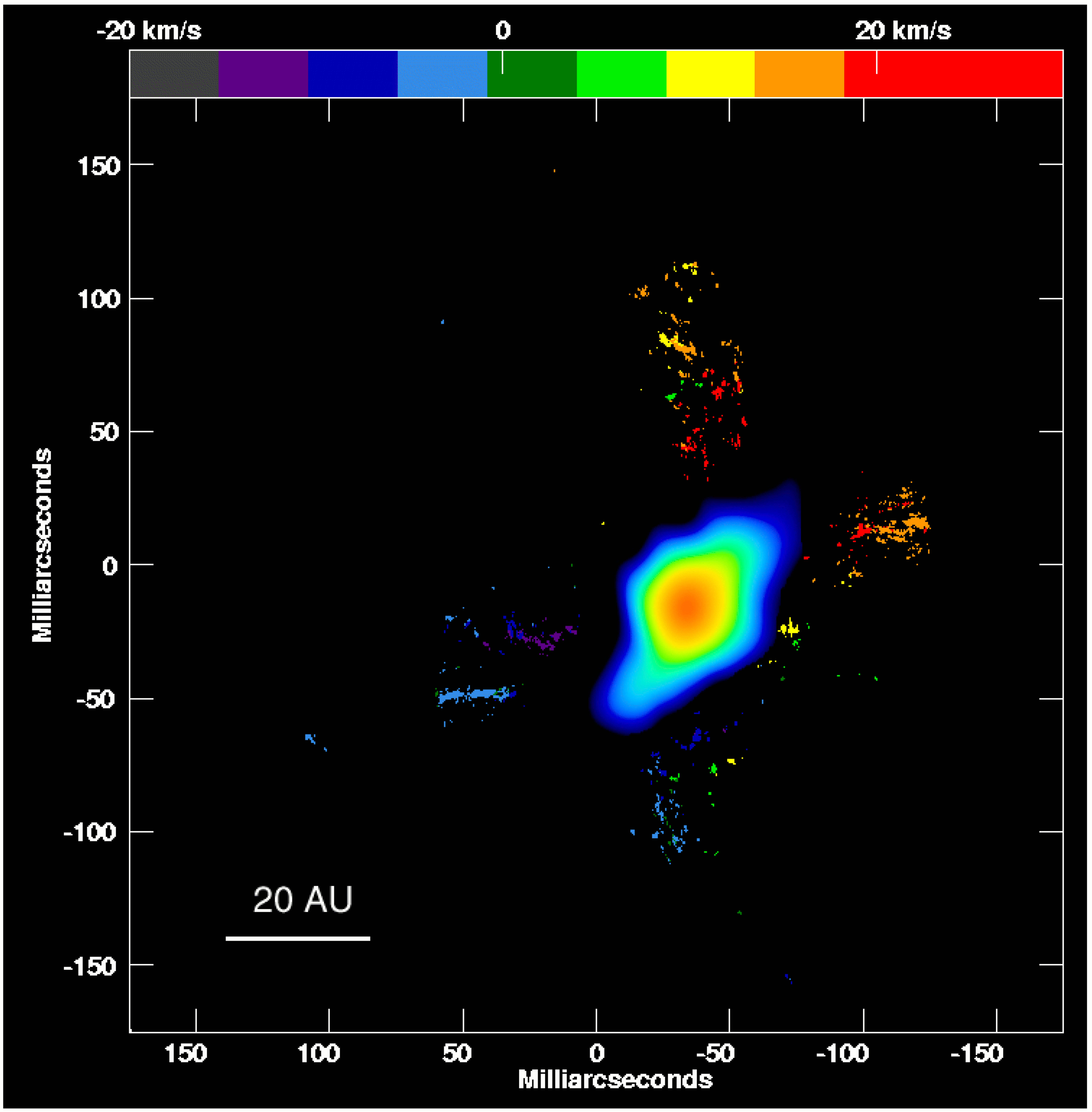}
\caption{Velocity field (in km/s) of the SiO $v$=1,2 $J$=1-0 maser emission in Orion Source I as observed
with the VLBA \citep{Matthews:2010p781}. The stellar velocity is $\sim$6 km s$^{-1}$. 
The bulk of the SiO emission is located within four
"arms" of an X-shaped pattern, while a $\sim$14 AU thick band with no SiO emission harbors the 7mm continuum source \citep{Reid:2007p9519, Goddi:2011p8577}. The continuum emission traces an ionized disk 
while the SiO masers probe material ablated from the surface of the disk in a  wind.}
\label{Obs_figure}
\end{figure}

Radio source I in Orion BN/KL  is the nearest example of a high-mass YSO. 
The close distance and exceptional richness in molecular maser tracers enabled the best detailed picture of  gas dynamics  in a massive YSO  on scales comparable with the Solar System. 
In particular, detailed imaging of  vibrationally-excited states SiO
masers with the Very Long Baseline Array (VLBA) enabled resolving the
surface of an edge-on disk as well as a wide-angle outflow in the
collimation and launch region at radii 10-100~AU
(\citealt{Matthews:2010p781}; Figure~\ref{Obs_figure}). Additionally,
at larger radii (100-1000 AU), the wide-angle wind appears to
collimate in a straight bipolar flow, as traced by H$_2$O and $\nu$=0 SiO
maser and thermal emission \citep{Plambeck:2009p10531, Greenhill:2012p10712}. Measurement of
the 3D velocity fields of SiO masers provides a clear evidence of rotation  in both the disk and the wind along the same axis. 
\cite{Cunningham:2005p8732}.
explored the importance of stellar winds from a luminous central source and proposed the idea that SiO masers in Source I may arise
from the interaction of a spherical wind with an infalling envelope. However, \cite{Matthews:2010p781} argued that such a model could not 
explain in detail the gas dynamics and physical conditions as revealed by SiO masers and alternately proposed
that the SiO masers arise from material
ablated from the surface of an accretion disk and discussed
qualitatively the applicability of various classes of disk-wind models
to Source I.
They find evidence to support that magnetic fields may play a role in driving and/or shaping the disk-wind from Source I, based on details of gas dynamics on small scales, like curved and helical trajectories traced by maser motions.  
Assuming a magnetic field is present, a magneto-centrifugally launched wind \citep[e.g.,][]{Blandford:1982p892,
  Konigl:2000p607, Fendt:2006p574} would be a natural candidate for powering a disk-wind from Source I. The latter thus  becomes an ideal laboratory to extend the MHD models from solar-like stars to  massive YSOs. 
 Recently, \cite{Vaidya:2011p8992} explored the interplay of
 radiation forces and magnetic fields to show that radiative line driving
 from luminous stars with masses $>30\,\msun$ does modify the dynamics
 of outflows. However, since the luminosity for a $10\,\msun$ binary Orion Source I would be less than a single $20\,\msun$, 
we adopt that model in the ideal MHD regime
 (neglecting radiative force) to numerically simulate the dynamics of
 the disk-wind inside 100 AU.

\section{Model Assumptions}
\label{modelA}
 \begin{figure} 
 \centering
 \includegraphics[width=1.0\columnwidth]{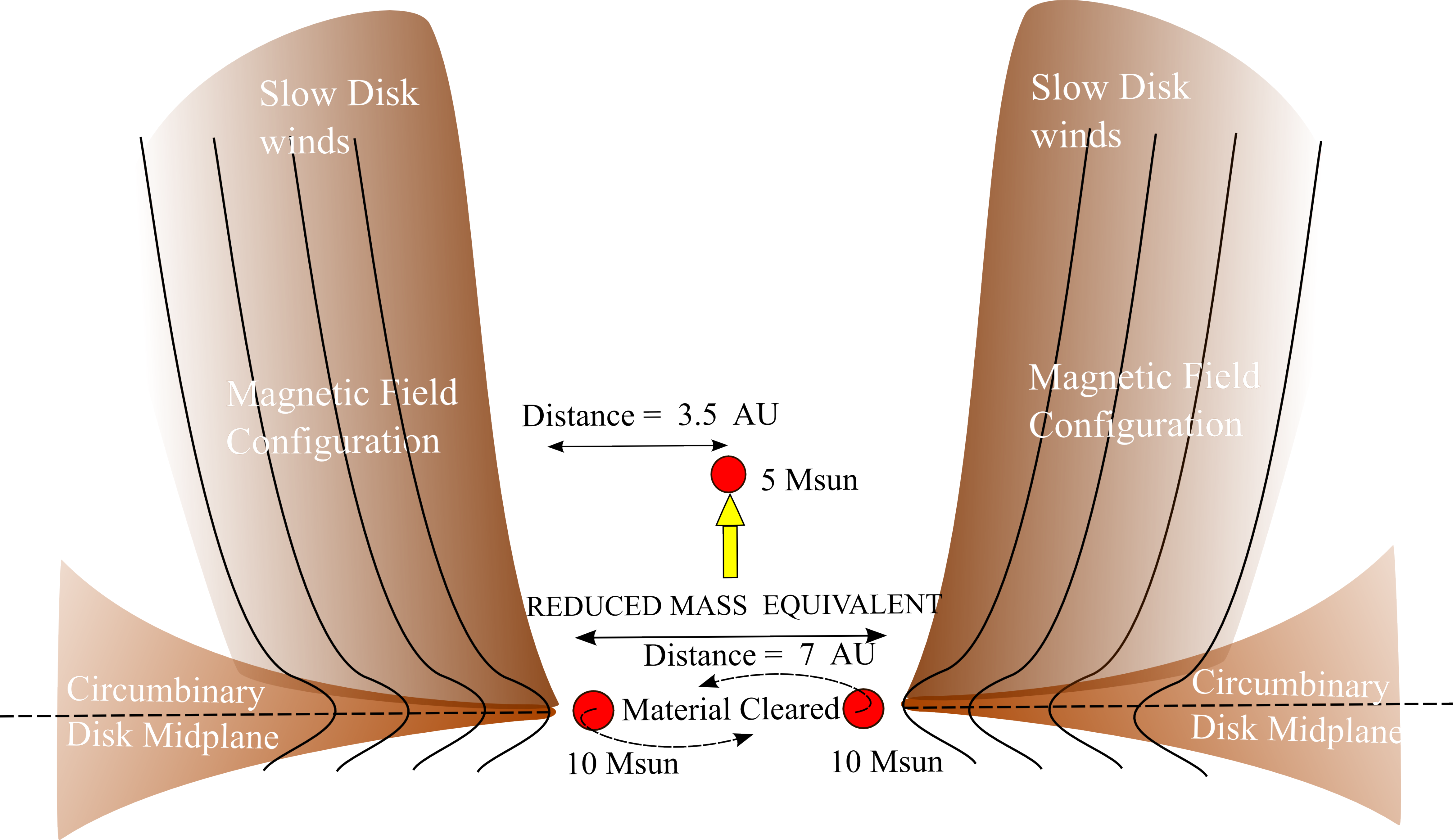}\\
 \caption{Cartoon Model for Source I.} 
 \label{SourceI_cartoon}
 \end{figure}
The initial conditions adopted to model the disk-wind from Orion Source I are reported in Table~\ref{ParaSummary} and are derived  from either observations and/or modeling of the system. 
Source I is believed to be an eccentric, hard  binary with a total
mass of $2\times10$ M$_{\odot}$, 
based on studies of the dynamics of YSOs and molecular gas in Orion BN/KL, as well as N-body simulations \citep{Goddi:2011p8577,Bally:2011p9419}. 
In particular, proper motion measurements of radio sources  provided strong evidence that a dynamical interaction occurred about 500~years ago between massive YSOs Source~I and BN  \citep[e. g.,][]{Goddi:2011p8577}.  
 N-body numerical simulations showed that the dynamical interaction
 between a binary of two 10\,$\msun$\,stars (Source I) and a single 10\,$\msun$\,star
  (BN) may lead to simultaneous ejection of both stars  and the BN/KL outflow
as well as binary hardening,  while preserving the original circumbinary disk  around Source I
 \citep{Bally:2011p9419, Moeckel:2011p8574}. 
Mechanical energy conservation (and virial theorem) implies an orbit semi-major axis $a_0=
 2-5$ AU, depending on the estimate of the kinetic energy of the
 runaway stars and the molecular gas flow \citep{Goddi:2011p8577,
   Bally:2011p9419}.  
For simplicity, we neglect the eccentricity  
  and assume that the system is an equal-mass circular
binary, comprised of two $10\,\msun$ stars  orbiting in a
circle of radius $a_0 = 3.5$\,AU.

A high-resolution study of H$\alpha$ and [O I]$\lambda$6300 line profiles in a pre-main-sequence spectroscopic binary has shown evidence of a bipolar jet as being launched by the whole binary  as opposed to each star individually  \citep{Mundt:2010p5}. Similarly, we assume  that the binary members  orbit inside the inner radius  of a truncated circumbinary disk and that  the wind  is launched at this radius.   
The cartoon figure representing our model assumption is shown in Fig.~\ref{SourceI_cartoon}.

To study the wind launching from Source I,
we perform axi-symmetric ideal MHD simulations in cylindrical
coordinates (r,$\phi$,z)  using the PLUTO code \citep{Mignone:2007p644}. 
We essentially prescribe a hydrostatic corona
(density and pressure) threaded with force-free magnetic field. 
An adiabatic equation of state, $P=\frac{\gamma-1}{\gamma}\rho^{\gamma}$, with $\gamma = 5/3$, relates the pressure with the density in the flow. 
For the central gravity, we treat the system of equal mass binary members as an equivalent $5\,\msun$ (reduced
mass) object at their barycenter. The accretion disk which is treated as a boundary forms the
launching base of the wind \cite[e.g.,][]{Ouyed:1997p634}
Further, we choose the wind launching point to be at a distance equal to the radius of the binary orbit,
i.e. $r_0$ = $a_0$, where the material is rotating with a
sub-Keplerian velocity $V_{\phi} = \chi v_0$. 
Further, the disk is considered to be
denser than the hydrostatic corona and this density contrast is
prescribed by $\delta$. In a realistic (thick and hot)
disk, the central gravity is balanced radially by contributions from thermal
pressure and centrifugal rotation. This can be ensured by
consistently deriving the sub-Keplerianity($\chi$) based on the choice
of density contrast. For example, smaller values of $\delta$ implies a
hotter (denser) disk and a larger contribution
of thermal pressure resulting in the disk to rotate with sub-Keplerian
speeds to maintain radial balance. We chose $\delta \sim 3$ as a 
reference value implying the underlying disk to rotate at a speed 0.8
times the Keplerian speed.

%
The magnetic field strength is prescribed by the value of plasma beta (the ratio of thermal pressure to
the magnetic pressure) at $r_0$, $\beta_0$. 
In order to put constraints on the magnetic field
strengths in Source I, we study the wind 
properties with varying $\beta_0$   (see Sect.~\ref{magfpara}). 
Finally, we assume the number density at $r_0$ to be  $n_0$ = $3 \times 10^{10} \rm cm^{-3}$. 
\cite{Reid:2007p9519} argued the  disk in Source I to be ionized 
with a typical electron density of $\sim 10^{7}\rm cm^{-3}$ (their "Model C"),  which corresponds to an
ionization fraction of $\sim 10^{-3}$.  

\begin{table}
\begin{center}
\caption{Parameters chosen for
  the reference run to study MHD driven outflow from Orion Source I. }
\begin{tabular}{ l l }
\hline\hline
\noalign{\smallskip}
Quantity & Reference Value \\
\hline
\noalign{\smallskip}
Masses of Binary[$m_{1}$, $m_{2}$]   & $10\,\msun, 10\,\msun$\\
\noalign{\smallskip}
Orbital Separation [2$a_0$] & 7\,AU\\
\noalign{\smallskip}
Reduced Stellar Mass [$\mu_{*}$]& $5\,\msun$\\
\noalign{\smallskip}
Wind Launching radius [$\rm{r}_0$] & 3.5\,AU\\
\noalign{\smallskip}
Keplerian Velocity at $\rm{r}_0$ [$\rm{v}_0$] & 35.7 km s$^{-1}$\\
\noalign{\smallskip}
Volume Density at $\rm{r}_0$ [$\rho_0$] & $5.0\times10^{-14}$
g\,cm$^{-3}$\\
\noalign{\smallskip}
Plasma Beta at $\rm{r}_0$ [$\beta_0$] & 1.0 \\
\noalign{\smallskip}
Density Contrast [$\delta$] & 2.85 \\
\noalign{\smallskip}
\hline
\noalign{\smallskip}
\end{tabular}\label{ParaSummary}
\end{center}
\end{table}

The flow is freely allowed to leave the outer boundaries (r,z = 50
$a_0$, 150 $a_0$) and symmetric boundary conditions are applied on the r=0 axis. 
The poloidal component of the magnetic field at the outer boundaries
is set to avoid artificial collimation by ensuring zero current
($\nabla \times \vec{B} = 0$; \citealt{Ustyugova:1999p655, Porth:2010p1041}).
 The critical boundary at z=0 mimics the inflow of gas and magnetic flux from the disk surface.
Here, we ensure the disk to be in pressure equilibrium with the
medium but with a higher density (i.e. colder than the medium). 
The angular speed of the magnetic field line in the disk is equal to the Keplerian
value and the poloidal velocity is updated each time with the value at
the grid cell just above the boundary. 
 Further, ideal MHD approximation permits to set the flow velocity along the poloidal magnetic field ($v_{\rm  p} \parallel B_{\rm p}$).
Thus, the mass flux at the disk boundary is not fixed but consistently derived by causal
interaction between the outflowing gas and the boundary
value. Detailed description of these boundary conditions are given in
\citet{Vaidya:2011p8992}.

\section{Results and Discussion}
\subsection{MHD driven wind profiles}

\begin{figure*} \centering
\includegraphics[width=1.9\columnwidth]{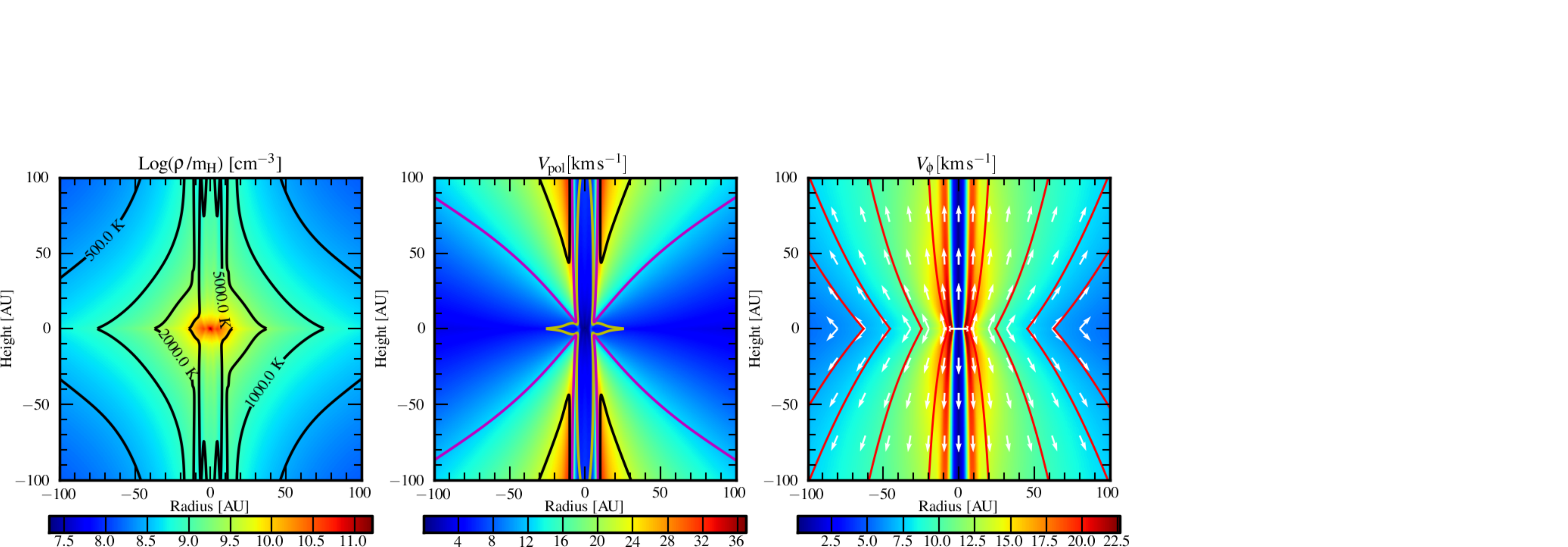}
\caption{Number density (left), poloidal velocity
  (middle) and azimuthal velocity (right) of the reference run in
  steady-state. Temperature contours (black) for T =
  500\,K, 1000\,K, 2000\,K and 5000\,K are overlaid in the left
  panel. The middle panel shows the positions of
  three MHD critical surfaces - magneto-sonic surface (yellow),
  Alfv$\acute{e}$n surface (magenta), magneto-fast surface (black). 
  White arrows in the right panel represents the velocity
  vectors which are parallel to magnetic field lines shown by solid
  red lines.} 
\label{phyvelref}
\end{figure*}

At the onset of our reference run, material at the base of the disk is ejected in vertical
direction along the magnetic field. This material propagates outwards
producing bow shocks in the ambient medium. 
As this shock propagates outside of the domain, the flow collimates due to magnetic forces
until a steady equilibrium is attained, beyond which no further
evolution of magnetic field lines takes place. 
The steady-state values of the density, poloidal and azimuthal velocities of the wind are shown in 
Figure~\ref{phyvelref}, along with the magnetic field lines (in red), the poloidal velocity versors (white
arrows), temperature contours and MHD critical surfaces. The magnetic
field lines ideally mark the morphology of the outflow. 
The left panel of Figure~\ref{phyvelref}  shows the steady-state
number density of the fluid, whose values range from 10$^{8}$\,cm$^{-3}$ to 10$^{11}$\,cm$^{-3}$. Overlaid on the density
map are the temperature contours, with values ranging from 5000\,K closer to the central source to 100\,K at large
distances. The high density and temperatures close to the axis reflect the fact that the initial hydrostatic conditions are fixed at the
boundary inside of the wind launching radius (this ensures numerical stability of the flow very close to the axis).
The middle panel of Figure~\ref{phyvelref} shows the  poloidal  wind velocity, whose values range from 8.0 to 32.0 km s$^{-1}$. 
The lines over-plotted show  three critical surfaces in a steady-state MHD flow,
viz. the magneto-sonic surface (yellow), the magneto-fast surface (black), 
and the Alfv$\acute{e}$n surface (red). 
The flow starts sub-magneto-sonically from the disk and is
accelerated to super fast speeds mainly via Lorentz forces. 
The Alfv$\acute{e}$n surface marks the locus
of points in the flow where ideally the kinetic energy of the flow is
converted to magnetic energy. This results in bending of the field
lines thereby generating a toroidal field and corresponding hoop stress
which collimates the flow. 
The right panel of Figure~\ref{phyvelref} finally shows the azimuthal component of the wind velocity, with values ranging from 10
km\,s$^{-1}$ at the outer edge to 25 km\,s$^{-1}$ close to the inner
wind launching point. 
The transfer of angular momentum from the wind to the medium may impart rotation to the latter, as observed in SiO masers (Sec.~\ref{siocomp}). 

\subsection{Comparison between MHD wind models and  wind properties observed from SiO masers}
\label{siocomp}
We propose that the SiO masers in Orion Source I
are excited as a moderately rotating disk-wind interacts with the ambient
medium via slow MHD shocks and transfer of angular momentum. 
Similar mechanisms have been successfully applied to explain the excitation of
$\rm H_{2}$O  masers \citep{Kaufman:1996p9607}.
Here, we qualitatively compare the steady-state properties of a
single-fluid, rotating MHD wind (launched from an underlying accretion disk)
with the observed physical and dynamical properties of the outflow molecular component probed by SiO masers in  Source I.
 
Three physical properties of the wind from the MHD modeling are consistent with SiO maser observations:
temperature/density profiles, morphology, and velocity field. 

  The conditions for optimum excitation of the SiO maser emission are  n(H$_2$)$\sim 10^{9} \rm cm^{-3}$ and T$\lesssim 2000\,\rm K$, respectively, as derived
 from a radiative transfer analysis by \citet{Goddi:2009p8571}. In our
 simulations, the regions close to the inner launching point (z,r
 $\lesssim$  15\,AU) have values of densities (10$^{10-11}$ cm$^{-3}$) and temperatures (2000-5000 K)  beyond the critical values required for SiO maser excitation
 (Fig.~\ref{phyvelref}, left panel). This is  consistent with the 14 AU thick "dark band" seen in
 Fig.~\ref{Obs_figure} where no masers are observed. The intermediate region (15\,AU $<$ z,r $<$
 60\,AU) has typical densities of $10^{9} \ \rm cm^{-3}$ and $T \sim 1000-2000\,K$ (Fig.~\ref{phyvelref}, left panel). Here, the wind can provide a
 strong IR background source for pumping the SiO masers in the shocks
 produced by its interaction with the ambient molecular medium. 

In the steady-state, the magnetic collimation and
 acceleration result in an open funnel shaped
 structure of field lines  (Fig.~\ref{phyvelref}). 
 Such a configuration of magnetic flux surfaces
 naturally results in a disk-wind with wide-angle bipolar morphology, which
 resembles the X-shaped wind structure around Source I revealed by SiO maser emission (Fig.~\ref{Obs_figure}). 

Regarding the velocity field, the dynamical properties of molecular outflows can be extrapolated from the single-fluid MHD wind by analyzing the momentum transfer \citep{Masson:1993p9661}. In this study, the density of the MHD driven wind  exceeds the mean density of its surrounding medium. Such dense and cold winds carry a large amount of momentum and impart a significant fraction of it to the ambient medium, resulting in the  shocked  material to have velocities of the same order as the  wind itself  \citep{Masson:1993p9661}, i.e. in the range of 10 - 30 km\,s$^{-1}$ (Fig.~\ref{phyvelref}, middle panel). 
Consistently, \cite{Matthews:2010p781} measured the 3D velocities of the molecular gas flows around 
Source I, to vary in the range from 5 km s$^{-1}$ to
25 km s$^{-1}$, with an average of 14 km s$^{-1}$. 
In addition, SiO masers show rotation seen as a velocity gradient in the
radial velocities of the four arms of the "X" tracing the disk-wind, 
which is consistent with the trend obtained for $V_{\phi}$ (Figure~\ref{phyvelref}, right panel). 

Therefore, the simulated MHD winds can drive molecular flows by interacting via
shocks with the ambient medium. 

\begin{figure} \centering
\includegraphics[width=0.9\columnwidth]{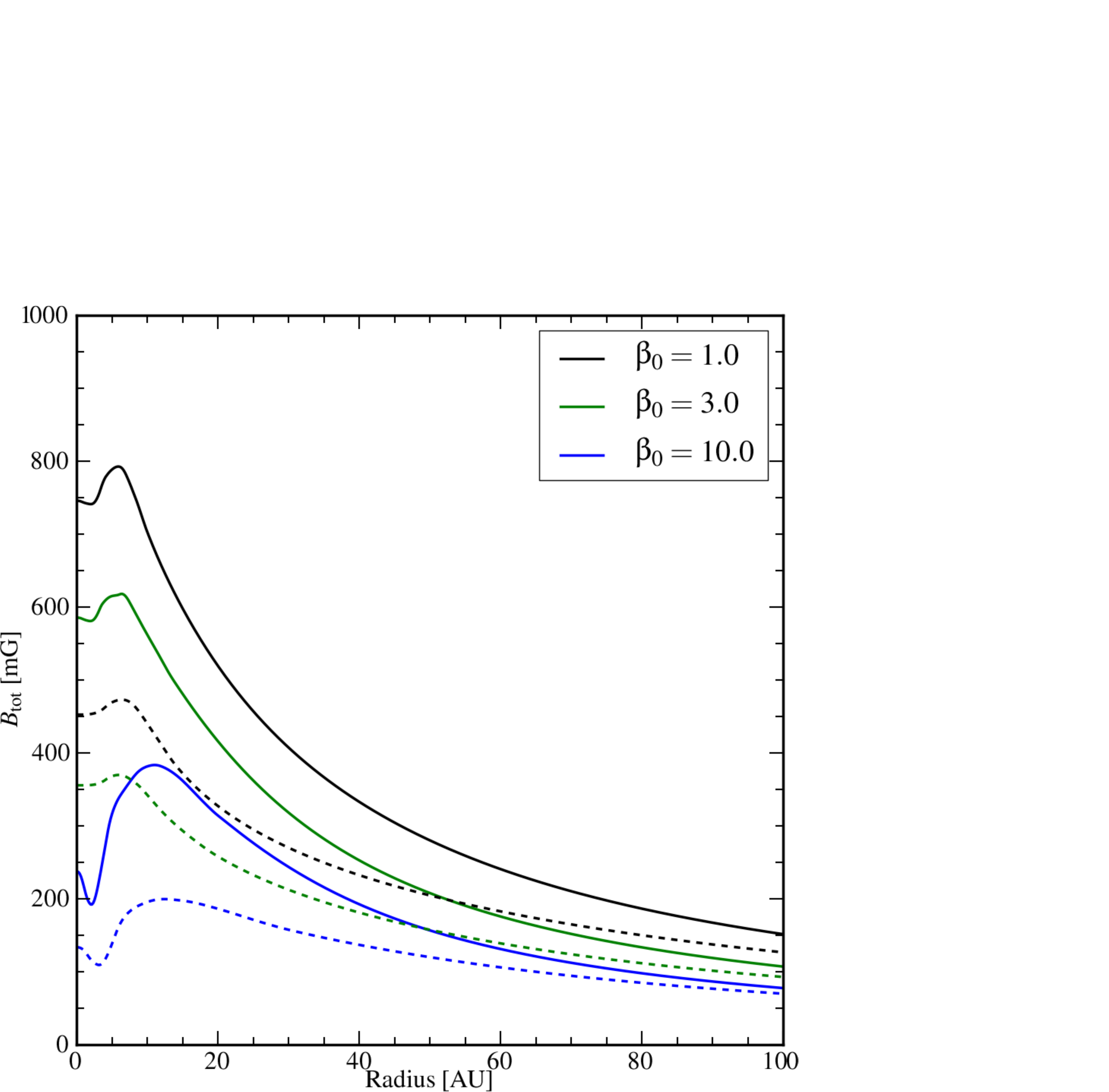}
\caption{Radial profiles of the magnetic field [in mG] at z =
  20\,AU [solid] and z = 60\,AU [dashed] for the reference run, obtained after the MHD wind has reached a steady-state.} 
\label{phyBtot}
\end{figure}


\subsection{Magnetic field strength}
\label{magfpara}
In our models, the magnetic field strength is governed by $\beta_0$.
We have carried out simulation runs with three different 
values of $\beta_0$ viz. 1.0, 3.0 and 10.0. 
Figure~\ref{phyBtot} 
shows the radial profiles of  magnetic field strength at z = 20\,AU
and z = 60\,AU for different values of $\beta_0$ and $\delta$=
2.85. 
We find that the magnetic field strength varies in the range 0.2--0.8 G (depending on $\beta_0$) and decreases
with radius and/or vertical height. 
From this study, we observe that the changes in density and
temperature are not significant with varying $\beta_0$ (e.g., $\Delta T \sim 100-200\,K$ when $\beta_0$ increases by ten fold). 
On the other hand, poloidal and azimuthal velocities decrease substantially with
increasing values of $\beta_0$. Since lower field strengths imply lower
Lorentz force, the flow is not efficiently accelerated resulting in
lower velocities. The typical velocities obtained from $\beta_0 > 3.0$
are not enough to explain the observed velocities in Source I. 
Therefore, we assume as a reference the magnetic field strength corresponding to
$\beta_0 = 1$, $\sim$0.35 G (at r, z =40\,AU, 20\,AU).
Assuming equipartition, \citet{Matthews:2010p781} quote a magnetic field strength of $\sim$0.3\,G for
Source I, based on kinematics of SiO masers. This value is consistent
with a measurement  of OH 1665 MHz masers across a $\sim 10^4$ AU region around Source I, which provided field strengths of 1.8 to 16.3 mG \citep{Cohen:2006p9565}. 
Since the OH masers are believed to arise from material with
$n(H_2)\sim 10^7$ cm$^{-3}$ \citep{Gray:1992p9570} and the magnetic 
field strength is expected to scale as the square root of the gas
density,  a simple extrapolation from the large scales (probed by OH masers) to the small scales (probed by SiO masers)  
gives field strengths of order the value predicted by the equipartition assumption.  
These empirical estimates  lie comfortably within the range predicted by our
model, but direct measurements via the Zeeman effect could provide a more reliable comparison  between models and observations.

\subsection{Model Limitations}
\label{limitations}

The main qualitative features of a steady MHD driven disk-wind
are consistent with the physical and dynamical properties of the molecular flow probed by SiO masers. 
A more quantitative analysis requires however further theoretical and observational studies. In this section, we list the 
limitations of the present model and the scope for future work.

(i) The assumption of treating the disk as a boundary allows us to follow wind dynamics till 
   a steady state is reached, but neglects the impact of  disk
  dynamics on wind propagation. A more realistic model should allow
   both disk and   outflow to  evolve simultaneously
 \cite[e.g.,][]{Casse:2002p1142, Sheikhnezami:2012p10641}. Also, we
 neglect any feedback on the disk-wind from a stellar wind (Source I has however  a rather modest $L_{\rm bol} \sim 10^{4}\,L_{\odot}$;
 \citealt{Gezari:1998p10535}).

(ii) In our ideal MHD model, we have assumed a single fluid MHD driven wind and have ignored effects due to diffusion of
  neutral molecules across the magnetic field. 
Though the regions
  close to the star are hot and potentially have a high
  degree of ionization, consistent with an ideal approximation,
  a more accurate model should incorporate effects due
  to diffusion between the neutral and ionized medium.
  Future simulations will relax the ideal MHD approximation and
   explore  effects of magnetic diffusivity.

(iii) We have ignored any feedback due to molecular chemistry. 
  Ideally, a direct comparison between MHD models and maser observations   would require a   detailed multi-fluid chemical model. Recently, conditions for molecular survival in MHD
disk-winds  were computed using chemical evolution of a self-similar
MHD wind model \citep{Panoglou:2012p10039}. 
However, such a chemical model
  with micro-physics of maser excitation is quite complex. As a follow-up, we plan to include non-equilibrium cooling and separately follow the evolution of
neutral and ionized gas fraction in the flow. 

(iv) Including  in the modeling the elements discussed above,  requires measurement of relevant physical quantities within the launching region ($\lesssim
  100$\,AU), i.e. ionization degree, molecular abundances, and
  magnetic field strengths.
High resolution and sensitive observations of  Hydrogen radio recombination lines (with JVLA and ALMA) will help us constrain the dynamical and physical properties of ionized gas. Further, radiative transfer analysis of rotational molecular lines (from ALMA) can constrain molecular gas properties and abundances. Finally, polarimetric observations of SiO masers (with VLBI) will enable to measure magnetic field strength and orientation at the disk/jet interface.

\section{Summary}
We have explored the plausibility of MHD driven wind
model to explain the disk-wind properties in Orion Source I as probed by SiO masers. In
particular, we have performed axisymmetric MHD wind launching
simulations treating Source I as an equal mass
binary and choosing initial parameters  derived from observations. 
We propose that the SiO masers are excited as the MHD driven wind interacts with the ambient
molecular medium in form of shocks. 
Four elements of consistency support this hypothesis:

\begin{enumerate}
\item
Our model consistently reproduces the bipolar morphology of the disk-wind traced by the SiO masers. 

\item
The poloidal and azimuthal velocities from the simulations are in the range of 10 -- 30 km s$^{-1}$, which
agree very well with the observed average 3D velocity of 14 km s$^{-1}$ and with the rotational component of SiO masers.

\item
The physical conditions in the modeled wind are consistent with the requirements for
pumping SiO masers. 

\item
We predict  field strengths of 0.2-0.8 G within 100 AU, in agreement
with empirical estimates from SiO masers.
\end{enumerate}

We conclude that the main  dynamical features of the SiO masers disk-wind in Orion Source I can be
 explained by the MHD  wind model. 
In a broader context,  Source I has provided a unique laboratory  
 to extend the standard MHD models from low-mass to high-mass YSOs, indicating that  in both regimes
similar physics may describe the transfer of matter and angular momentum between disks and outflows. 

\section*{Acknowledgments}
We thank Christian Fendt and Lynn Matthews for their valuable
comments. B.V thanks the STFC for funding.

\bibliographystyle{mn2e}

\begin{thebibliography}{}

\bibitem[\protect\citeauthoryear{{Bally}, {Cunningham}, {Moeckel}, {Burton},
  {Smith}, {Frank} \& {Nordlund}}{{Bally} et~al.}{2011}]{Bally:2011p9419}
{Bally} J.,  {Cunningham} N.~J.,  {Moeckel} N.,  {Burton} M.~G.,  {Smith} N.,
  {Frank} A.,    {Nordlund} A.,  2011, \apj, 727, 113

\bibitem[\protect\citeauthoryear{{Blandford} \& {Payne}}{{Blandford} \&
  {Payne}}{1982}]{Blandford:1982p892}
{Blandford} R.~D.,  {Payne} D.~G.,  1982, \mnras, 199, 883

\bibitem[\protect\citeauthoryear{{Casse} \& {Keppens}}{{Casse} \&
  {Keppens}}{2002}]{Casse:2002p1142}
{Casse} F.,  {Keppens} R.,  2002, \apj, 581, 988

\bibitem[\protect\citeauthoryear{{Cohen}, {Gasiprong}, {Meaburn} \&
  {Graham}}{{Cohen} et~al.}{2006}]{Cohen:2006p9565}
{Cohen} R.~J.,  {Gasiprong} N.,  {Meaburn} J.,    {Graham} M.~F.,  2006,
  \mnras, 367, 541

\bibitem[\protect\citeauthoryear{{Cunningham}, {Frank} \&
  {Hartmann}}{{Cunningham} et~al.}{2005}]{Cunningham:2005p8732}
{Cunningham} A.,  {Frank} A.,    {Hartmann} L.,  2005, \apj, 631, 1010

\bibitem[\protect\citeauthoryear{{Fendt}}{{Fendt}}{2006}]{Fendt:2006p574}
{Fendt} C.,  2006, \apj, 651, 272

\bibitem[\protect\citeauthoryear{{Gezari}, {Backman} \& {Werner}}{{Gezari}
  et~al.}{1998}]{Gezari:1998p10535}
{Gezari} D.~Y.,  {Backman} D.~E.,    {Werner} M.~W.,  1998, \apj, 509, 283

\bibitem[\protect\citeauthoryear{{Goddi}, {Greenhill}, {Chandler}, {Humphreys},
  {Matthews} \& {et. al.}}{{Goddi} et~al.}{2009}]{Goddi:2009p8571}
{Goddi} C.,  {Greenhill} L.~J.,  {Chandler} C.~J.,  {Humphreys} E.~M.~L.,
  {Matthews} L.~D.,    {et. al.} 2009, \apj, 698, 1165

\bibitem[\protect\citeauthoryear{{Goddi}, {Humphreys}, {Greenhill}, {Chandler}
  \& {Matthews}}{{Goddi} et~al.}{2011}]{Goddi:2011p8577}
{Goddi} C.,  {Humphreys} E.~M.~L.,  {Greenhill} L.~J.,  {Chandler} C.~J.,
  {Matthews} L.~D.,  2011, \apj, 728, 15

\bibitem[\protect\citeauthoryear{{Gray}, {Field} \& {Doel}}{{Gray}
  et~al.}{1992}]{Gray:1992p9570}
{Gray} M.~D.,  {Field} D.,    {Doel} R.~C.,  1992, \aap, 262, 555

\bibitem[\protect\citeauthoryear{{Greenhill}, {Goddi}, {Chandler}, {Humphreys}
  \& {Matthews}}{{Greenhill} et~al.}{2012}]{Greenhill:2012p10712}
{Greenhill} L.~J.,  {Goddi} C.,  {Chandler} C.~J.,  {Humphreys} E.~M.~L.,
  {Matthews} L.~D.,  2012, in IAU Symposium Vol.~287, {}.
pp 166--170

\bibitem[\protect\citeauthoryear{{Kaufman} \& {Neufeld}}{{Kaufman} \&
  {Neufeld}}{1996}]{Kaufman:1996p9607}
{Kaufman} M.~J.,  {Neufeld} D.~A.,  1996, \apj, 456, 250

\bibitem[\protect\citeauthoryear{{Konigl} \& {Pudritz}}{{Konigl} \&
  {Pudritz}}{2000}]{Konigl:2000p607}
{Konigl} A.,  {Pudritz} R.~E.,  2000, Protostars and Planets IV, p.~759

\bibitem[\protect\citeauthoryear{{Masson} \& {Chernin}}{{Masson} \&
  {Chernin}}{1993}]{Masson:1993p9661}
{Masson} C.~R.,  {Chernin} L.~M.,  1993, \apj, 414, 230

\bibitem[\protect\citeauthoryear{{Matthews}, {Greenhill}, {Goddi}, {Chandler},
  {Humphreys} \& {Kunz}}{{Matthews} et~al.}{2010}]{Matthews:2010p781}
{Matthews} L.~D.,  {Greenhill} L.~J.,  {Goddi} C.,  {Chandler} C.~J.,
  {Humphreys} E.~M.~L.,    {Kunz} M.~W.,  2010, \apj, 708, 80

\bibitem[\protect\citeauthoryear{{Mignone}, {Bodo}, {Massaglia}, {Matsakos},
  {Tesileanu}, {Zanni} \& {Ferrari}}{{Mignone} et~al.}{2007}]{Mignone:2007p644}
{Mignone} A.,  {Bodo} G.,  {Massaglia} S.,  {Matsakos} T.,  {Tesileanu} O.,
  {Zanni} C.,    {Ferrari} A.,  2007, \apjs, 170, 228

\bibitem[\protect\citeauthoryear{{Moeckel} \& {Goddi}}{{Moeckel} \&
  {Goddi}}{2012}]{Moeckel:2011p8574}
{Moeckel} N.,  {Goddi} C.,  2012, \mnras, 419, 1390

\bibitem[\protect\citeauthoryear{{Mundt}, {Hamilton}, {Herbst}, {Johns-Krull}
  \& {Winn}}{{Mundt} et~al.}{2010}]{Mundt:2010p5}
{Mundt} R.,  {Hamilton} C.~M.,  {Herbst} W.,  {Johns-Krull} C.~M.,    {Winn}
  J.~N.,  2010, \apjl, 708, L5

\bibitem[\protect\citeauthoryear{{Ouyed} \& {Pudritz}}{{Ouyed} \&
  {Pudritz}}{1997}]{Ouyed:1997p634}
{Ouyed} R.,  {Pudritz} R.~E.,  1997, \apj, 482, 712

\bibitem[\protect\citeauthoryear{{Panoglou}, {Cabrit}, {Pineau Des For{\^e}ts},
  {Garcia}, {Ferreira} \& {Casse}}{{Panoglou}
  et~al.}{2012}]{Panoglou:2012p10039}
{Panoglou} D.,  {Cabrit} S.,  {Pineau Des For{\^e}ts} G.,  {Garcia} P.~J.~V.,
  {Ferreira} J.,    {Casse} F.,  2012, \aap, 538, A2

\bibitem[\protect\citeauthoryear{{Plambeck}, {Wright}, {Friedel}, {Widicus
  Weaver}, {Bolatto}, {Pound}, {Woody}, {Lamb} \& {Scott}}{{Plambeck}
  et~al.}{2009}]{Plambeck:2009p10531}
{Plambeck} R.~L.,  {Wright} M.~C.~H.,  {Friedel} D.~N.,  {Widicus Weaver}
  S.~L.,  {Bolatto} A.~D.,  {Pound} M.~W.,  {Woody} D.~P.,  {Lamb} J.~W.,
  {Scott} S.~L.,  2009, \apjl, 704, L25

\bibitem[\protect\citeauthoryear{{Porth} \& {Fendt}}{{Porth} \&
  {Fendt}}{2010}]{Porth:2010p1041}
{Porth} O.,  {Fendt} C.,  2010, \apj, 709, 1100

\bibitem[\protect\citeauthoryear{{Reid}, {Menten}, {Greenhill} \&
  {Chandler}}{{Reid} et~al.}{2007}]{Reid:2007p9519}
{Reid} M.~J.,  {Menten} K.~M.,  {Greenhill} L.~J.,    {Chandler} C.~J.,  2007,
  \apj, 664, 950

\bibitem[\protect\citeauthoryear{{Sheikhnezami}, {Fendt}, {Porth}, {Vaidya} \&
  {Ghanbari}}{{Sheikhnezami} et~al.}{2012}]{Sheikhnezami:2012p10641}
{Sheikhnezami} S.,  {Fendt} C.,  {Porth} O.,  {Vaidya} B.,    {Ghanbari} J.,
  2012, \apj, 757, 65

\bibitem[\protect\citeauthoryear{{Ustyugova}, {Koldoba}, {Romanova},
  {Chechetkin} \& {Lovelace}}{{Ustyugova} et~al.}{1999}]{Ustyugova:1999p655}
{Ustyugova} G.~V.,  {Koldoba} A.~V.,  {Romanova} M.~M.,  {Chechetkin} V.~M.,
  {Lovelace} R.~V.~E.,  1999, \apj, 516, 221

\bibitem[\protect\citeauthoryear{{Vaidya}, {Fendt}, {Beuther} \& {et.
  al.}}{{Vaidya} et~al.}{2011}]{Vaidya:2011p8992}
{Vaidya} B.,  {Fendt} C.,  {Beuther} H.,    {et. al.} 2011, \apj, 742, 56

\end{thebibliography}

\label{lastpage}
\end{document}